\documentclass[aps,pra,twocolumn,showpacs]{revtex4}
\usepackage[latin1]{inputenc}
\usepackage{graphicx,amsmath,amsfonts,amssymb}
\usepackage{color}
\usepackage{graphicx}
\usepackage{epstopdf}
\newcommand{\be}{\begin{equation}}
\newcommand{\ee}{\end{equation}}

\newcommand{\ket}[1]{\mbox{$ | #1 \rangle $}}
\newcommand{\bra}[1]{\mbox{$ \langle #1 | $}}

\begin{document}

\title{Irrealism from fringe visibility in matter-wave double-slit interference with initial contractive states}
\author{F. R. Lustosa$^{1,2}$}
\author{P. R. Dieguez$^{1}$}\email{dieguez.p@ufabc.edu.br}
\author{I. G. da Paz$^{2}$}\email{irismarpaz@ufpi.edu.br}

\affiliation{$^1$ Centro de Ci\^{e}ncias Naturais e Humanas, Universidade Federal
do ABC, Avenida dos Estados 5001, 09210-580, Santo Andr\'{e}, SP, Brazil}

\affiliation{$^2$ Departamento de F\'{\i}sica, Universidade Federal
do Piau\'{\i}, Campus Ministro Petr\^{o}nio Portela, CEP 64049-550,
Teresina, PI, Brazil}

\begin{abstract}

The elements of reality coined by Einstein, Podoslky, and Rosen promoted a series of fundamental discussions involving the notion of quantum correlations and physical realism. The
superposition principle applied in the double-slit experiment with
matter waves highlights the need for a critical review of the
adoption of physical realism in the quantum realm. In this work, we employ a measure of physical irrealism and consider an initial contractive state in the double-slit setup for which position and momentum variables of a single particle are initially correlated. 
We investigate how the behavior of the irrealism can help us to obtain information about the interference pattern, wavelike, and particle-like properties in the double-slit
setup with matter waves. We find that there is a time of propagation
that minimizes the irrealism, and around this point the state at the
detection screen is squeezed in position and momentum in
comparison with the standard Gaussian superposition. Interestingly, we
show that the maximum visibility and the number of interference
fringes are related to the minimum of the irrealism. Moreover, we demonstrate a monotonic relation between the irrealism and visibility around the time of minimum. Then, we
argue how to use these results to indirectly measure the irrealism for position variable from the fringe visibility.

\end{abstract}


\maketitle

\section{Introduction}

The Loophole-free violation of Bell's inequalities leaves no doubt
that the classical deterministic notion of an objective reality
calls for a critical
review~\cite{bell64,hensen15,giustina15,shalm15}. One could say that
this idea starts with the celebrated work \cite{EPR35} of Einstein, Podolsky, and Rosen (EPR), where
they introduced a sufficient condition to describe an element of
physical reality. EPR wondered whether it was possible to ascribe
realistic properties to a quantum system in the absence of
measurements, bringing up a formal definition to probe physical
realism in the context of quantum theory. Employing also a necessary
criterion for completeness of a physical theory, they argued that
quantum mechanics was incomplete since it would allow simultaneous
elements of reality for incompatible observables. EPR's criterion is
related to certainly predict the value of some physical property,
without disturbing it, then assuming also the condition of causality
in space-time~\cite{dieguez18}.

Bohr's approach to that was in terms of his complementarity
principle~\cite{bohr35}, which says that the elements of reality of
incompatible observables cannot be established in the same
experiment, but only through mutually excluding experimental
arrangements. The notion of wave particle duality, which arises in
the complementarity principle, highlights the counterintuitive
character of quantum mechanics. According to that, the same physical
system can exhibit either a particle-like or a wave-like
behavior \cite{Feynman,Bohr} depending on the specific design of a
certain experiment. The nontrivial result of that principle is well
illustrated with the so-called Wheeler's delayed choice gedanken
experiment. On the other hand, we know today that with the addition
of a quantum-controlled device of the delayed-choice experiment, it
is possible to challenge the complementarity principle by allowing
the same experimental apparatus to measure wave-to-particle
transitions~\cite{Ionicioiu11, Auccaise12, Peruzzo12, Kaiser12}.

Matter-waves quantum interference, another notable aspect of nature in
which massive particles exhibit spatial delocalization, completely
challenging our classical intuition about physical realism, is also
a subject of intense research given its importance to the
foundations of quantum theory.  Experiments reveling
wave-particle duality in the double-slit were performed by
M\"{o}llenstedt and J\"{o}sson for electrons \cite{Jonsson}, by
Zeilinger et al. for neutrons \cite{Zeilinger1}, by Carnal and
Mlynek for atoms \cite{Carnal}, using diffraction gratings by
Sch\"{o}llkopf and Toennies for small molecules \cite{Toennies}, by
Zeilinger et al. for macromolecules \cite{Zeilinger2}, and electron
double-slit diffraction has been experimentally observed in
\cite{Bach}. Moreover, the Einstein-Bohr debate about the
wave-particle duality in the ``floating" double-slit gedanken
experiment, has recently been explored in \cite{liu2015}. Using
molecules as slits, this provides an experimental proof and
theoretical support showing that a Doppler marker eliminates the
interference pattern, in corroboration with Bohr's complementary
principle \cite{liu2015}. Interestingly, this consideration goes
against the logic initially advocated by EPR and emphasizes the role
of correlations generated in the experimental
configuration~\cite{bilobran15,angelo15, dieguez18} as a fundamental mechanism to the emergence of physical realism.

Conceptually different from quantum correlations between two systems
or two different Hilbert spaces of a single system (for example,
spin and position degrees of freedom), position-momentum
correlations are correlations that indicate a dependence
between the position and the momentum of a single particle. In the
case of a simple Gaussian or minimum-uncertainty wavepacket solution
for the Schr\"{o}dinger equation for a free particle, the
position-momentum correlations at $t=0$ are zero but they appear for
later times \cite{Bohm,Saxon}. On the other hand, more complex
states such as squeezed states or a linear combination of Gaussian
states can exhibit initial correlations, i.e., correlations that do
not depend on the time evolution
\cite{Robinett,Riahi,Dodonov,Campos}. It was shown that the
existence of position-momentum correlations is related with the
phases of the wave function \cite{Bohm}. The position-momentum
correlations can also be used to gain information about other relevant physical quantities. For example, qualitative changes were shown in the interference
pattern as a function of the increase in the position-momentum
correlations \cite{Carol}. The Gouy phase matter waves are
directly related to the position-momentum correlations, as studied
in Ref.\cite{Paz1}. A relation was observed
between the position-momentum correlations and the formation of
above-threshold ionization (ATI) spectra in the electron-ion
scattering in strong laser fields \cite{Kull}. More recently, it was
shown that the maximum of the position-momentum correlations is
related with the minimum number of interference fringes in the
double-slit experiment \cite{solano}.

In this work, we use the facts that the measure of physical
(ir)realism introduced by Bilobran and Angelo~\cite{bilobran15} for
discrete-spectrum observables is quantitative, operational, and was
further extended for continuous variables in~\cite{freire19}, thus
allowing us to make formal connections of this quantity, and other
measures such as wave-like and particle-like properties in the
double-slit experiment with matter waves modeled by a
contractive state~\cite{Yuen}, to verify how the evolution of the
fundamental and entropic uncertainties affects these measures. This
paper is structured as follows. In Sec. II, we introduce and
briefly discuss the main properties of the measure of the degree of
irrealism developed in \cite{bilobran15}. In Sec. III, we model
the double-slit experiment with matter waves considering an initially
contractive Gaussian wavepacket, which propagates during the time
$t$ from the source to the double-slit and during the time $\tau$
from the double-slit to the screen. We calculate the wave functions
for the passage through each slit using the Green's function for the
free particle to obtain the position-momentum correlations, the
fundamental and  entropic uncertainties, and the irrealism for this
system that is a linear combination of the states that passed
through each slit. We show that these quantities are minima at the
same propagation time from the source to the double-slit. On the
other hand, the position-momentum correlations and the fundamental
uncertainties have a common point of maximum that does not coincide
with the maximum irrealism. In Sec. IV, the relative intensity,
visibility, and predictability are analyzed in terms of the minimum
and maximum of the irrealism. For the viewpoint of comparison, we
also analyze these quantities for the maximum of the
position-momentum correlations. Here, it was proposed to indirectly
measure the irrealism from the fringe visibility around the time of
minimum. In Sec. V we present our concluding remarks.

\section{Irrealism and quantum theory}
\label{irrealism}

There is not a single formalism to discuss the concept of realism in
quantum theory, and we should not be surprised about it if we
remember that there is also not a unified interpretation of the
consequences of the basic axioms of quantum theory. Indeed, the
puzzle of defining realism in quantum theory invariably touches
other foundational questions, such as  the measurement problem and
wave-particle duality. For example, in a traditional double-slit
setup, does the quantum system pass through both slits
simultaneously, acting somehow like a definite wave, or we should
interpret it as a state of fundamental local indefiniteness in that
region of space-time? What has to be considered as a physical
premiss to probe physical realism in this context?

When questioning whether it was possible to attribute realism to
entangled quantum systems, EPR was based both on the completeness of a
physical theory and also on a sufficient condition that would define
realistic aspects. According to EPR, the necessary condition assumed
for a physical theory to be considered complete is that every
element of physical reality must have a correspondence in the
theory~\cite{EPR35}. To conduct their analysis in what concerns the
elements of physical reality, they define the following sufficient
condition:  ``If, without disturbing the system in any way, we can
predict with certainty the value of a physical quantity, then there
is an element of physical reality corresponding to this quantity".
When applied to spins, this condition can be stated as follows.
Suppose the singlet state,
$\ket{\phi}_{\cal{AB}}=\frac{1}{\sqrt{2}}(\ket{0}_{\cal{A}}\ket{1}_{\cal{B}}-\ket{1}_{\cal{A}}\ket{0}_{\cal{B}})$.
After the formation of this entangled pair, it is clear that under a
measurement on one of the systems, the spin in the same direction of
the other can always be anticipated with certainty since this
correlation is preserved in all bases. Then, according to EPR there
must be an element of physical reality corresponding to this
physical quantity. Note that this conclusion is independent of the
spatial distance between the systems, thus it is possible to assume
a strong version of relativistic causality~\cite{vaidman96}, that
is, no space-like event could perturb the local results. A deep
consequence of that argument, together with the fact that we can
measure the spin along any direction and the correlation is the
same, is that the fact of an exact prediction on a distant system
invites us to interpret this as the mere manifestation of a
pre-existing value of this physical property, since both systems did
not interact while the measurement was carried out. This implies
that incompatible observables also have their elements of reality
defined independently. Since quantum theory generally does not give us
this value for observables that do not commute, EPR concluded that
although correct, it gives an incomplete description of reality,
being necessary as discussed by EPR, the so-called hidden variables
that would complete the theory. As we know today, no set of hidden
local variables is capable of reproducing all the results of
experiments with entangled states. Furthermore, as EPR attributes
realism only to properties that can be predicted with certainty,
such a criterion does not include the class of mixed states that
have their probabilities linked to a purely subjective ignorance, as
in the case of epistemic states.

To discuss the definition of physical realism in the
double-slit experiment with matter waves, we review in this section,
a quantifier introduced by Bilobran and Angelo
~\cite{bilobran15}, which put forward an operational scheme to
assess elements of reality of discrete-spectrum observables in
quantum mechanics and generalizes EPR's definition. The main idea of
this measure is constructed under the premise that a measurement
establishes the reality of an observable even if we do not have
access to the result of this measurement, thus indicating realism
also for an epistemic state, which is a state that has only
subjective ignorance.

This can be formally stated with the following procedure.  Bilobran and Angelo \cite{bilobran15}
consider a preparation $\rho$ that acts in $\cal{H_A\otimes H_B}$ submitted to a
protocol of unrevealed measurements (also known as non-selective measurements) of a generic observable
$A=\sum_aaA_a$, with projectors $A_a=\ket{a}\bra{a}$, acting on
$\cal{H_A}$. Since the outcome of the measurement is considered unrevealed in
this protocol, the resulting state is the average over all possible results,
\be
\label{reality}
\Phi_A(\rho)\equiv\sum_{a}\left(A_a\otimes
1_{\mathcal{B}}\right)\,\rho\,\left(A_a\otimes
1_{\mathcal{B}}\right)=\sum_ap_aA_a\otimes\rho_{\mathcal{B}|a}, \ee where
$\rho_{\mathcal{B}|a}=\bra{a}\rho\ket{a}/p_a$ and $p_a=\text{Tr}[(A_a\otimes
1_{\mathcal{B}})\rho]$.
Bilobran and Angelo propose to take $\Phi_A(\rho)$ as a
state of reality for $A$ and $\rho=\Phi_A(\rho)$ as a formal
criterion of reality. Note that this premise of realism also agrees with the EPR criterion,
since eigenstate preparations are elements of reality for
some observable, but this also attempts to generalize EPR in a sense
that it also quantifies the degree of realism for mixed states. With that, by employing the relative
entropy $S(\rho||\sigma)\equiv\text{Tr}[\rho(\ln{\rho}-\ln{\sigma})]$, they compute the degree of
irrealism of the observable $A$ given the preparation $\rho$ as
\be \label{frakI}
\mathfrak{I}(A|\rho)\equiv S(\rho||\Phi_A(\rho))=S(\Phi_A(\rho))-S(\rho),
\ee
where $S(\rho)\equiv-\text{Tr}(\rho\ln{\rho})$ is the von Neumann
entropy. Note that, this quantifier is non-negative and vanishes if
and only if $\rho=\Phi_A(\rho)$, thus allowing us to interpret it as
an ``entropic distance" between the state $\rho$ and the state that obeys
realism for this observable $\Phi_A(\rho)$. As discussed in
\cite{bilobran15,freire19}, although one could use some other norm,
the use of the ``entropic metric" allows one to relate this measure
with other quantities of quantum information theory. For example,
the above formula can be decomposed as
\be \label{LIQC}
\mathfrak{I}(A|\rho)=\mathfrak{I}(A|\rho_{\cal{A}})+D_A(\rho),
\ee
where $D_A(\rho)=I_{\cal{A:B}}(\rho)-I_{\cal{A:B}}(\Phi_A(\rho))$
stands for the nonminimized version of the one-way quantum discord
(see Refs.~\cite{bilobran15,dieguez18} for further details). So, the
irrealism of $A$ is the sum of local coherence (that is, the
coherence of $A$ given the reduced state $\rho_{\cal{A}}$) with
quantum correlations associated with measurements of $A$. Then, for
single-partite states ($\rho$  which acts in $\cal{H_A}$) or
uncorrelated bipartite states, the irrealism reduces to the amount
of quantum coherence relative to the $A$-Basis
$\mathfrak{I}(A|\rho_{\cal{A}})$~\cite{baumgratz14}. This observation highlights the
contextual character of irrealism, since it is an observable and
also a state-dependent quantity.

This approach gives a prominent role to the notion of information.
Indeed, by employing a model of measurement called
monitoring~\cite{dieguez18,dieguez2018}, it was deduced that there is a formal
connection between information and reality in quantum mechanics,
developing a complementarity relation to these
concepts~\cite{dieguez18}.  By now, this measure has proven relevant
in scenarios involving coherence \cite{angelo15},
nonlocality~\cite{gomes18,gomes19,fucci19}, weak
reality~\cite{dieguez18}, which also has an experimentally
verification with photonic weak measurements~\cite{mancino18},
realism-based entropic uncertainty relations~\cite{rudnicki18},
random quantum walk~\cite{orthey19}, Hardy's
paradox~\cite{engelbert20}, and more recently, from the point of
view of a generalized resource theory of information~\cite{costa20}.
Nevertheless, all of these works are exclusively applied to discrete
spectrum observables. To fill this gap, Freire and Angelo presented
a framework in~\cite{freire19}, which calculates the degree of
realism associated with continuous variables such as position and
momentum by explicitly presenting a formalism through which one can
quantify the degree of irrealism associated with a continuous
variable for a given quantum state, by showing how to consistently
discretize the position and momentum variables in terms of
operational resolutions of the measurement apparatus. With that,
they implemented  an operational notion of projective measurement
and a criterion of reality for these quantities. Interestingly, they
introduced a quantifier for the degree of irrealism of a discretized
continuous variable, which, when applied to pure states, exhibits an
uncertainty relation to the conjugated pair position-momentum, as
\be \mathfrak{I}(Q|\rho)+\mathfrak{I}(P|\rho)\geq ln(2\pi e), \ee
meaning that quantum mechanics, equipped with Heisenberg's
uncertainty relation, prevents classical realism for conjugated
quantities~\cite{freire19}.

In this work, we propose to apply these ideas to
formally analyze how position and momentum elements of reality
behave in situations in which massive particles produce an interference
pattern due to a coherent superposition in the double-slit setup, in
order to interpret this state as assuming a purely physical
uncertainty regarding its reality. So, in what follows we discuss
how we model our double-slit setup and discuss how we quantify the
degree of irrealism in such a context.

\section{Irrealism in the double-slit experiment}

\begin{figure}[htp]
\centering
\includegraphics[width=7.0 cm]{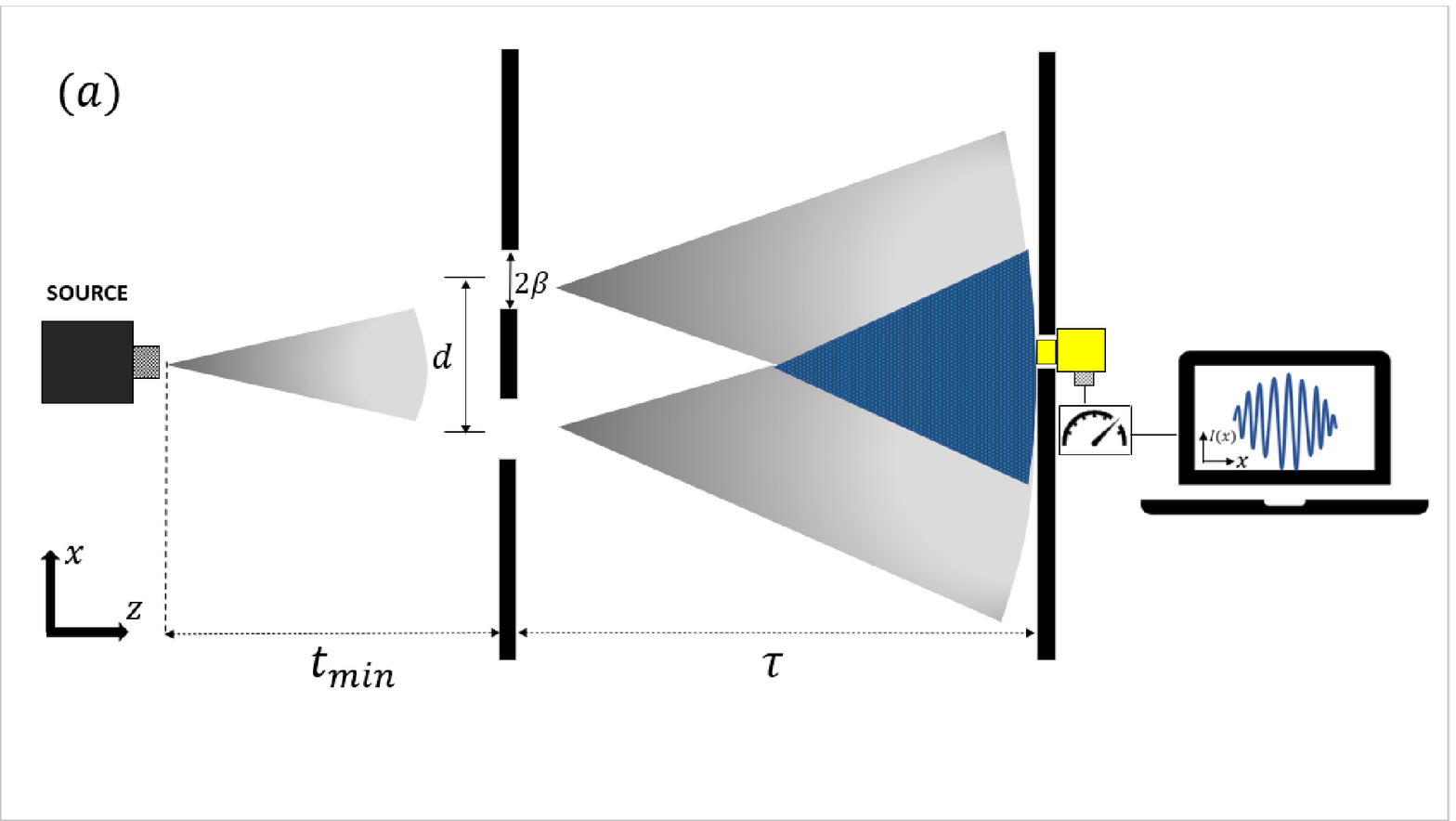}
\includegraphics[width=7.0 cm]{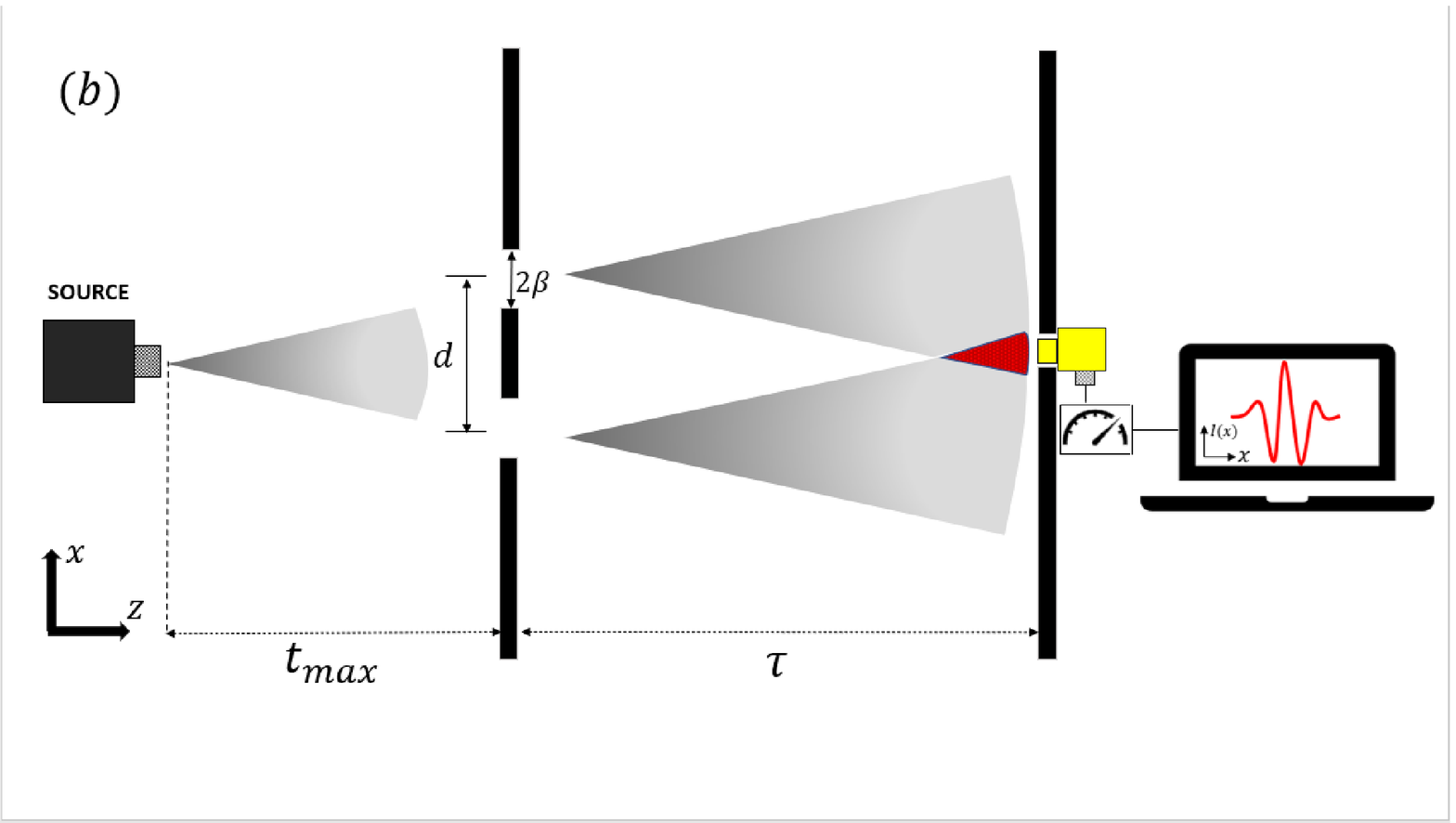}
\caption{Sketch of the double-slit experiment. A correlated Gaussian
wavepacket of transverse width $\sigma_{0}$ propagates during a time
$t$ before attaining the double-slit and during a time $\tau$ from
the double-slit to the screen. The slit transmission functions are
taken to be Gaussian of width $\beta$ and separated by a distance
$d$. In (a) the initial wavepacket propagates a time $t_{\textsl{min}}$ from
the source to the double-slit which produces at the detection screen
the maximum region of overlap, interference fringes, and visibility.
In (b) the initial wavepacket propagates a time $t_{\textsl{max}}$ from the
source to the double-slit which corresponds to a small region of
overlap with a small number of interference fringes and
visibility.}\label{Figure1}
\end{figure}

In this section, we model the double-slit experiment as follows.
Before reaching the double-slit setup, we consider  that a coherent
Gaussian wave packet of initial
width $\sigma_{0}$ correlated in position and momentum propagates during a time $t$ before arriving at a
double-slit that divides it into two Gaussian wavepackets. These
initial correlations are measured by a parameter $\gamma$, such
that, for $\gamma\neq 0$, the state is spread in momentum but acquires
a portion of correlations such that the Robertson-Schr\"odinger
uncertainty relation attains the minimum value of $\hbar^{2}/4$. It has
been shown that this state is not squeezed in terms of the
conventional position and momentum operators but it exhibits
squeezing for the generalized quadrature operators defined through
the conventional operators by a rotation in the phase space
\cite{IGP2020}. After the double-slit, the two wavepackets propagate
during a time $\tau$ until they reach the detection screen, where
they are recombined and the interference pattern is observed as a
function of the transverse coordinate $x$. Here, we consider 
a 
one-dimensional problem such that the momentum in the $Oz$ direction is
very high and produces a very small wavelength and negligible
quantum effects. Then, we treat this direction as classical, and the
quantum effects are observed in the $Ox$ direction. As a consequence
of the free propagation, which decouples the $x$, $y$, and $z$
dimensions for a given longitudinal location, we can write $z=v_{z}t$
for the $Oz$ direction. The position and momentum of the particle
will be correlated, and such correlations, as well as the fundamental
and entropic uncertainties, will be changed by the evolution and the
parameter $\gamma$.  This model is presented in 
Fig.\ref{Figure1}
together with illustrations of the behave that will be found in the
results. In Fig.\ref{Figure1}(a) the initial wavepacket propagates
a time $t_{\textsl{min}}$ from the source to the double-slit, which produces
at the detection screen the maximum region of overlap, interference
fringes, and visibility. As we will see in the results, this
propagation time corresponds to the minimum position-momentum
correlations, fundamental and entropic uncertainties, and irrealism.
In Fig.\ref{Figure1}(b), the initial wavepacket propagates a time
$t_{\textsl{max}}$ from the source to the double-slit, which produces at the
detection screen a small region of overlap with a small number of
interference fringes and visibility. As we will see later on, this
propagation time corresponds to the maximum position-momentum
correlations and fundamental uncertainties, which are different from the
time for the maximum irrealism.

The wavefunction at the time when the wave passes through the upper
slit $(+)$ or the lower slit $(-)$ is given by \cite{Carol}

\begin{eqnarray}
\psi_{\pm}(x,t,\tau)&=&\int_{-\infty}^{\infty}
dx_j\int_{-\infty}^{\infty} dx_iG_2(x,t+\tau;x_j,t)\nonumber\\
&\times&F(x_j\mp
d/2)G_1(x_j,t;x_i,0)\psi_0(x_i) ,
\end{eqnarray}
where the kernels $G_{1}$ and $G_{2}$
are the free propagators for the particle,
\begin{equation}
G_1(x_j,t;x_i,0)= \sqrt{\frac{m}{2\pi i \hbar t}} \exp
\left[\frac{im(x_j-x_i)^2}{2 \hbar t} \right],    
\end{equation}
\begin{equation}
G_2(x,t+ \tau;x_j,t)= \sqrt{\frac{m}{2\pi i \hbar \tau}} \exp
\left[\frac{im(x-x_j)^2}{2 \hbar \tau} \right]  ,  
\end{equation}
the function $F(x_{j}\mp d/2)$ describes the double-slit transmission functions, which are
taken to be Gaussian of width $\beta$ separated by a distance $d$,
\begin{equation}
F(x_j\mp d/2)= \frac{1}{\sqrt{\beta \sqrt{\pi}}}  \exp
\left[-\frac{im(x_j \mp d/2)^2}{2 {\beta}^2} \right]  ,  
\end{equation}
and considering the initial wave function as
\begin{equation}
\psi_0(x_i)= \frac{1}{\sqrt{\sigma_{0} \sqrt{\pi}}}  \exp
\left[-\frac{{x^2_i}}{{2\sigma^2_0}} + \frac{i \gamma
x^2_i}{2\sigma^2_0}\right]. 
\end{equation}
Here, we consider that $\sigma_{0}$ is the transverse width of the
initial wavepacket, $m$ is the mass of the particle, $t$ ($\tau$) is
the time of flight from the first slit (double-slit) to the
double-slit (screen). The parameter $\gamma$ ensures that the
initial state is correlated in position and momentum. In fact, we
obtain for the initial state $\psi_0(x_i)$ the following results for
the uncertainties in position and in momentum,
$\sigma_{xx}=\sqrt{\langle \hat{x}^2\rangle - \langle
\hat{x}\rangle^2}=\sigma_{0}/\sqrt{2}$ and $\sigma_{pp}=\sqrt{\langle
\hat{p}^2\rangle - \langle
\hat{p}\rangle^2}=\sqrt{1+\gamma^{2}}\hbar/\sqrt{2}\sigma_{0}$, and
for the position-momentum correlations $\sigma_{xp}=\frac{\langle
\hat{x}\hat{p}+\hat{p}\hat{x}\rangle}{2}-\langle
\hat{x}\rangle\langle\hat{p}\rangle=\hbar \gamma/2$. For $\gamma=0$,
we have the simple uncorrelated Gaussian wavepacket and for
$\gamma<0$, we have a contractive state~\cite{Yuen}. As we can observe the
Robertson-Schrödinger uncertainty relation
$\sigma_{xx}^{2}\sigma_{pp}^{2}-\sigma_{xp}^2$ attains the minimum
value $\hbar^2/4$ independent of $\gamma$.

To obtain analytic expressions for the fundamental
uncertainties, position-momentum correlations, relative intensity,
visibility, and predictability in the screen of detection, we use a
Gaussian transmission function instead of a top-hat transmission
function because a Gaussian transmission function represents a good
approximation to the experimental feasibility and also because it is
mathematically simpler to treat than a top-hat transmission
function. The corresponding wavefunction for the propagation through
the upper slit was previously obtained in Ref. \cite{oziel}, and it
is given by
\begin{eqnarray}\label{eqwave}
\psi_{+}(x,t,\tau)&=&\frac{1}{\sqrt{B_{\gamma}\sqrt{\pi}}}\exp
\Bigg[-\frac{(x-D_{\gamma}/2)^2}{2B_{\gamma}^2} \Bigg]{}
                           \nonumber\\
&&{}\times \exp\Bigg[\frac{imx^2}{2\hbar R_{\gamma}}
-i\Delta_{\gamma} x +i\theta_{\gamma} +i\mu_{\gamma} \Bigg],
\end{eqnarray}
where the wave-function parameters and their physical meanings are
discussed as follows. The classical optics analogous of
the radius of curvature of the wavefronts $R_{\gamma}$ for the
propagation through one slit is
\begin{equation}
R_{\gamma} (t,\tau)= \tau \frac{ \left(\frac{1}{\beta^2} +
\frac{1}{b_{\gamma}^2} \right)^2 + \frac{m^2}{\hbar^2}
\left(\frac{1}{ \tau} + \frac{1}{r}\right)^2}{ \frac{1}{\beta^4} +
\frac{C}{\sigma^4_0 (t^2 + \tau^2_0 + 2 \tau_0 t \gamma + t^2
\gamma^2)}},
\end{equation}
with $C=\left[ \tau^2_0 + \frac{t \tau^2_0}{\tau} + \tau^2_0
\gamma^2 + \frac{\tau^3_0 \gamma}{\tau} + \frac{t \tau^2_0
\gamma^2}{\tau} + \frac{2\tau^2_0 \sigma^2_0}{\beta^{2}} \right]$,
and $\tau_{0}=m\sigma_{0}^{2}/\hbar$ is one intrinsic time scale (the
half of the Ehrenfest time) which essentially corresponds to the
time during which a distance of the order of the wavepacket
extension is traversed with a speed corresponding to the dispersion
in velocity \cite{Roman}. It is viewed as a characteristic time for
the ``aging" of the initial state \cite{Carol,solano}, since it is a
time from which the evolved state acquires properties completely
different from the initial state. The wavepacket width $B_{\gamma}$
for the propagation through one slit is
\begin{equation}
B_{\gamma}^2(t,\tau)= \frac{ \left(\frac{1}{\beta^2} +
\frac{1}{b_{\gamma}^2} \right)^2 + \frac{m^2}{\hbar^2}
\left(\frac{1}{ \tau} + \frac{1}{r_{\gamma}}\right)^2}{
(\frac{m}{\hbar \tau})^2 \left(\frac{1}{\beta^2} +
\frac{1}{b_{\gamma}^2} \right)},
\end{equation}
$\Delta_{\gamma}$ is a time dependent wavenumber
\begin{equation}
\Delta_{\gamma}(t,\tau)=\frac{\tau \sigma^2_0 d}{2 \tau_0 \beta^2
B_{\gamma}^2},
\end{equation}
$D_{\gamma}$ is the separation between the
wavepackets produced in the double-slit
\begin{equation}
D_{\gamma}(t,\tau)=d
\frac{\left(1+{\frac{\tau}{r_{\gamma}}}\right)}{\left( 1+
\frac{\beta^2}{b_{\gamma}^2}\right)},
\end{equation}
$b_{\gamma}$ is the wavepacket width for the free propagation
\begin{equation}
b_{\gamma}(t)= \frac{\sigma_0}{\tau_0} \left[{ t^2 + \tau^2_0 + 2 t
\tau_0 \gamma+ t^2 \gamma^2 } \right]^\frac{1}{2},
\end{equation}
$r_{\gamma}$ is the classical optics analogous of the
radius of curvature of the wavefronts for the free propagation
\begin{equation}
r_{\gamma}(t)= \frac{ \left(  t^2 + \tau^2_0 + 2 t \tau_0 \gamma +
t^2 \gamma^2 \right) } { \left[  t \left( 1 +  \gamma^2 \right) +
\gamma\tau_0 \right] },
\end{equation}
and $\mu_{\gamma}$ is the Gouy phase for the
propagation through one slit
\begin{equation}
\mu_{\gamma}(t,\tau)= - \frac{1}{2} \arctan \left[ \frac{ t + \tau
\left( 1 + \frac{\sigma^2_0}{\beta^2} + \frac{t \hbar \gamma }{ m
\beta^2 } \right) } { \tau_0 \left( 1 - \frac{t \tau
\sigma^2_0}{\tau_0 \beta^2} \right) + \gamma \left( t + \tau
\right)} \right].
\end{equation}
Note that $\mu_{\gamma}$ and $\theta_{\gamma}$,
\begin{equation}
\theta_{\gamma}(t,\tau)= \frac{m d^2 \left( \frac{1}{\tau} +
\frac{1}{r_{\gamma}}\right)}{ 8 \hbar \beta^4 \left[ \left( \frac{1
}{\beta^2} + \frac{1}{b_{\gamma}^2}\right)^2 + \frac{m^2}{\hbar^2}
\left( \frac{1}{\tau} + \frac{1}{r_{\gamma}}\right)^2 \right ] },
\end{equation} are time
dependent phases and they are relevant only if the slits have
different widths. Also, to obtain the expressions for the wave function
$\psi_{-}(x,t,\tau)$ passing through the lower slit, we just have to
substitute the parameter $d$ by $-d$ in the expressions
corresponding to the wave passing through the upper slit. Different from the results obtained in
Ref. \cite{solano}, all the parameters above are changed by the
correlation parameter $\gamma$.

Now we are in a position to discuss how we evaluate the irrealism for
the wavefunction at the detection screen. First, we note that for
single partite pure states, it follows that $S(\rho)=0$, and the von
Neumann entropy of this state when the unread measurement
map is applied reduces to $S(\Phi_{Q(P)}(\rho))=H_{Q(P)}$, where $H_{Q(P)}$ is
the Shannon entropy associated with probability distributions for
the variables $q(p)$ (see \cite{freire19} for more details). To
construct the Shannon distribution of the discretized continuous variable
$x(p)$, 
we follow the main idea of Refs. \cite{birula84,partovi83} to write the
discretized probability distribution of a position measurement $p_m$
in terms of experimental resolutions for position measurements of
$\delta q$. However, since the expressions used in ~\cite{birula84,partovi83} are known to be ill-behaved in the limit of large coarse-graining for some states~\cite{rudnicki11}, here we take a slightly different strategy discussed in~\cite{rudnicki11}, which corrects this problem by introducing a symmetrical limit to compute the discretization procedure as \be p_{m}=\int_{(m-1/2)\delta q}^{(m+1/2)\delta q}
dx\;\rho(x,t,\tau,\gamma), \ee where
$\rho(x,t,\tau,\gamma)=\psi(x,t,\tau,\gamma)\times\psi^{*}(x,t,\tau,\gamma)$
is the probability density in position, and
\begin{equation}
\psi(x,t,\tau,\gamma)= \frac{\psi_{+} (x,t,\tau) + \psi_{-}
(x,t,\tau)} {\sqrt{2 + 2 \exp \left[ -  \frac{D_{\gamma}^2}{4
B_{\gamma}^2}  - \Delta_{\gamma}^2 B_{\gamma}^2
\right]}}, %
\label{PsiNorma}
\end{equation}
is the normalized wavefunction in the screen of detection.

Note that we made a discretization of the probability distribution
in terms of the experimental resolution $\delta q$, which is assumed
to be a constant. Moreover, with the introduction of the above symmetrical limit, which realizes that one shall investigate the limit of large experimental accuracies~\cite{rudnicki11}, we guarantee two important aspects of the irrealism measure. In their seminal paper, Bilobran and Angelo conceived it as an operational measure by adopting a measurement-based premise, thus linking the irrealism with the capabilities of the measurement apparatus. Furthermore, it was also shown in~\cite{dieguez18} that the irrealism is intimately related to the concept of information in quantum theory. So, it is expected that the irrealism goes to zero in the limit of very inaccurate experimental resolutions (see \cite{freire19} for more details), in the same way that very imprecise performed measurements tell us nothing about the information of quantum states.
With that, we can calculate the Shannon entropy as
\begin{equation}
H_{Q}(t,\tau,\gamma)=-\sum^{\infty}_{m=-\infty}{p_{m}\ln p_{m}}, \label{psix}
\end{equation}
and now it is straightforward to calculate the corresponding  degree of
irrealism for the position of the state (\ref{PsiNorma}), explicitly
by
\begin{eqnarray}
\mathfrak{I}(Q|\rho) = H_{Q}(t,\tau,\gamma)=-
\sum^{\infty}_{m=-\infty}{p_{m}\ln p_{m}}. \label{Irx}
\end{eqnarray}
For the momentum in terms of the wavevector $k$, we follow the same
strategy to write
\begin{eqnarray}
\mathfrak{I}(P|\rho) = H_{P}(t,\tau,\gamma)=-
\sum^{\infty}_{n=-\infty}{p_{n}\ln p_{n}}, \label{Irp}
\end{eqnarray}
with \be p_{n}=\int_{(n-1/2)\delta k}^{(n+1/2)\delta k}
dk\;\tilde{\rho}(k,t,\tau,\gamma), \ee where
$\tilde{\rho}(k,t,\tau,\gamma)$ is obtained with the corresponding
Fourier transform of $\psi(x,t,\tau,\gamma)$, and $p_{n}$ is the
discretized probability for momentum measurements with resolution
$\delta k$.

\subsection{Fundamental uncertainties relations}

In what follows, we calculate the fundamental uncertainties in
position and momentum, the position-momentum correlations, and the
generalized Robertson-Schr\"odinger uncertainty relation in the
detection screen. The fundamental uncertainties are important to establish the limit in which experimental resolutions would not be able to access quantum information.  We will show that the fundamental uncertainties and
the position-momentum correlations have a minimum and a maximum
common point as a function of the propagation time $t$ for a
negative value of the correlation parameter $\gamma$ (contractive
state). For positive or null values of $\gamma$ these quantities
have only a maximum, no minimum. On the other hand,
the Robertson-Schr\"odinger uncertainty relation has the same common
point of minimum for a contractive state, but it does not have a
maximum independent of the value and signal of $\gamma$.

For the normalized wavefunction Eq.(\ref{PsiNorma}), we calculate
the fundamental uncertainties in position and momentum, and the
position-momentum correlations, and we obtain, respectively,

\begin{equation}
\sigma^2_{xx} (t, \tau)= \frac{B_{\gamma}^2}{2} + \frac{D_{\gamma}^2
- 4 \Delta_{\gamma}^2 B_{\gamma}^4 \exp \left[ - \frac{D_{\gamma}^2
}{4 B_{\gamma}^2 } - \Delta_{\gamma}^2 B_{\gamma}^2 \right] }{ 4 + 4
\exp \left[ - \frac{D_{\gamma}^2 }{4 B_{\gamma}^2 } -
\Delta_{\gamma}^2 B_{\gamma}^2 \right] },
\end{equation}

\begin{eqnarray}
\frac{ \sigma^2_{pp} (t, \tau)}{ \hslash^2 } &=& \left(  \frac{1}{ 2
B_{\gamma}^2 } + \frac{m^2 B_{\gamma}^2 }{ 2 \hslash^2 R_{\gamma}^2
} \right) + \frac{ (\frac{m D_{\gamma} }{\hslash R_{\gamma}} - 2
\Delta_{\gamma})^2 }{ 4 + 4 \exp \left[ - \frac{D_{\gamma}^2 }{4
B_{\gamma}^2 } - \Delta_{\gamma}^2 B_{\gamma}^2 \right] }\nonumber\\
&-& \frac{\left[ \frac{D_{\gamma}^2}{ B_{\gamma}^4 } + 2
\Delta_{\gamma} \left( \Delta_{\gamma} + \frac{m D_{\gamma} }{
\hslash R_{\gamma}}\right) \right] }{1 + \exp \left[ -
\frac{D_{\gamma}^2 }{4 B_{\gamma}^2 } + \Delta_{\gamma}^2
B_{\gamma}^2 \right] },
\end{eqnarray}

and

\begin{eqnarray}
\sigma_{xp}(t, \tau)&=& \frac{m B_{\gamma}^2}{2 R_{\gamma}} + \frac{
\left( m D_{\gamma}^2/R_{\gamma} \right)}{4  + 4 \exp \left[-
\frac{D_{\gamma}^2}{4B_{\gamma}^2} - \Delta_{\gamma}^2 B_{\gamma}^2
\right]} - \frac{\hbar \Delta_{\gamma} D_{\gamma} }{2}\nonumber\\
&-& \frac{\left(   m \Delta_{\gamma}^2 B_{\gamma}^4 /R \right)}{1+
\exp \left[ \frac{D_{\gamma}^2}{4 B_{\gamma}^2} + \Delta_{\gamma}^2
B_{\gamma}^2 \right]}. \label{PM_C}
\end{eqnarray}

The determinant of the covariance matrix is the generalized
Robertson-Schr\"odinger uncertainty relation, and it is given by
\begin{equation}
D_{C}=\sigma_{xx}^{2}\sigma_{pp}^{2}-\sigma_{xp}^{2}.
\end{equation}

\begin{figure}[htp]
\centering
\includegraphics[width=4.5 cm]{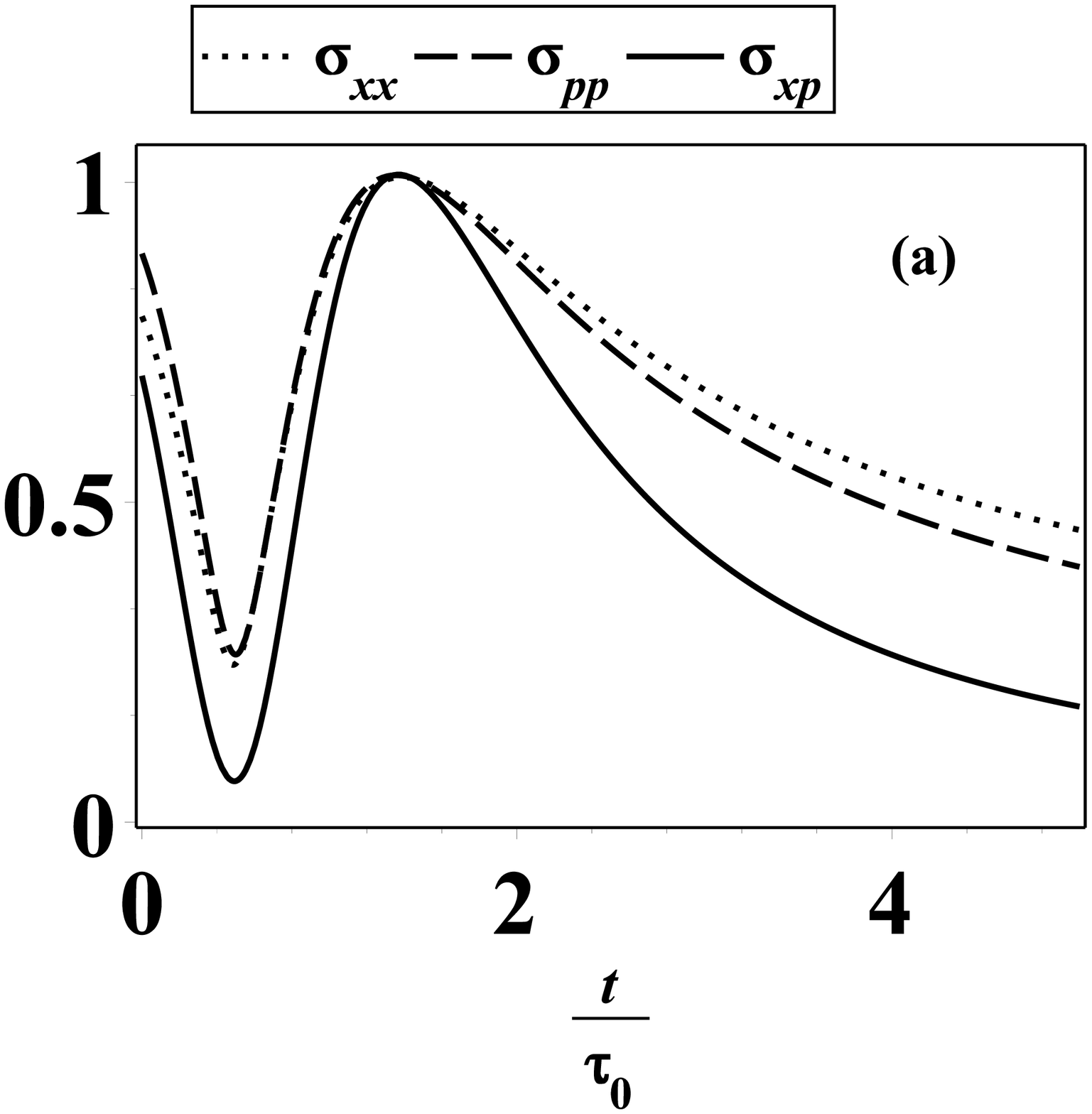}
\includegraphics[width=4.0 cm]{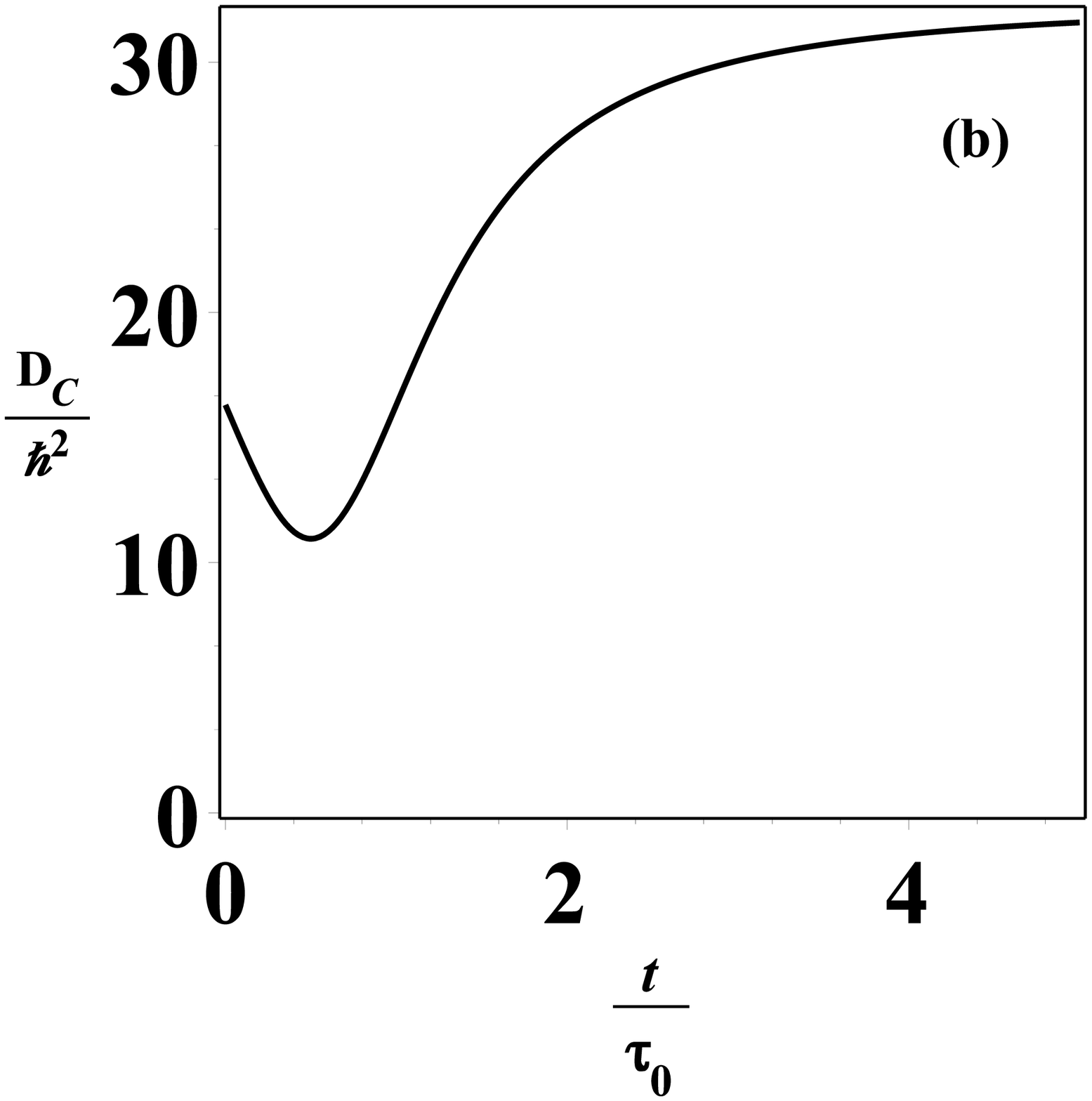}
\caption{(a) Uncertainties in position, in momentum and
position-momentum correlations and (b) Robertson-Schr\"odinger
uncertainty as a function of $t/\tau_{0}$ for $\tau=18\tau_{0}$ and
$\gamma=-1.0$. There is a common point of minimum for
$t_{\textsl{min}}\approx0.49\tau_{0}$ and a point of maximum for
$t_{\textsl{max}}\approx1.36\tau_{0}$. (b) The Robertson-Schr\"odinger
uncertainty has the same point of minimum but it does not have a
point of maximum. } \label{sigmas}
\end{figure}

In the following, we plot the curves for the uncertainties in
position, in momentum, the position-momentum correlations, and the
Robertson-Schr\"odinger uncertainty as a function of the time
$t/\tau_{0}$ for neutrons. The reason to consider neutrons relies on
their experimental feasibility, which is most close to our model for
interference with completely coherent matter waves, although we
still have a loss of coherence as discussed in Ref.\cite{Sanz}. We
adopt the following parameters: mass
$m=1.67\times10^{-27}\;\mathrm{kg}$, initial width of the packet
$\sigma_{0}=7.8\;\mathrm{\mu m}$ (which corresponds to the effective
width of $2\sqrt{2}\sigma_{0}\approx22\;\mathrm{\mu m}$), slit width
$\beta=7.8\;\mathrm{\mu m}$, the separation between the slits
$d=125\;\mathrm{\mu m}$, and de Broglie wavelength
$\lambda=2\;\mathrm{nm}$. These same parameters were used previously
in double-slit experiments with neutrons by Zeilinger et al.
\cite{Zeilinger1}. In Fig.\ref{sigmas}(a) we show the plot of the
uncertainties in position, momentum, and the position-momentum
correlations. In Fig.\ref{sigmas}(b) we show the
Robertson-Schr\"odinger uncertainty as a function of $t/\tau_{0}$
for $\tau=18\tau_{0}$, and for an initial contractive state
$\gamma=-1.0$. We use arbitrary normalization constants to
have the three curves at the same plot in Fig.\ref{sigmas}(a).
The dotted line corresponds to the uncertainty in position,
the dashed line corresponds to the uncertainty in momentum, and
the solid line corresponds to the position-momentum
correlations. As we can observe, these quantities have common points
of minimum and maximum that we calculate and obtain, respectively, as
$t_{\textsl{min}}\approx0.49\tau_{0}$ and $t_{\textsl{max}}\approx1.36\tau_{0}$. On
the other hand, the Robertson-Schr\"odinger uncertainty has the same
point of minimum but it does not have a point of maximum, which is a
consequence of how the position-momentum correlations evolve with
time.

A very interesting result is that despite the fact that the difference for the
times of maximum and minimum is smaller than unity in terms of
$\tau_{0}$, these values of time produce interference and fringe
visibility that are completely distinct, as we will see later on.

\subsection{Irrealism and fundamental uncertainties}

Following the formalism put forward in~\cite{bilobran15,dieguez18,freire19}, we present our results concerning the degree of irrealism in position and momentum variables for different experimental resolutions applied to our interferometer model. In particular, the fundamental uncertainties play an important role in the experimental resolution of the apparatus employed to probe the irrealism for continuous variables observables such as position and momentum. As long as the apparatus resolution is such that $\delta q < \sqrt{2\pi e}\sigma_{xx}$, we have positive values for the irrealism in position and momentum variables. Despite the dependence on the experimental resolution, we will show that we have the same time evolution for each different experimental resolution. Another important limit for the irrealism measure, discussed in greater detail in~\cite{freire19}, emerges when $\delta q \geq \sqrt{2\pi e}\sigma_{xx}$, since we have $\Im(Q|\rho)=0$ for this low-resolution regime. On the other hand, note that it is perfectly legitimate to explore the continuous limit of the Shannon entropy by following, for example, the procedure given in~\cite{Coles}, $h(x):=lim_{\delta q \longrightarrow 0}[H (x ) + ln \; \delta q]$, to compute the limit, which produces $H_{Q}=\ln\sqrt{2\pi e}\sigma_{xx}$ for Gaussian states, and $H_{Q}\leq \ln\sqrt{2\pi e}\sigma_{xx}$ for general states. However, here we focus on the operational meaning of irrealism coined in~\cite{bilobran15}, and we present our results in terms of experimentally feasible states $|k\delta q\rangle$, with $k \in \textbf{Z}$ for the role of the apparatus in our double-slit interferometer setup, instead of thinking in the idealized eigenstates $|x\rangle$. Also, one can follow the formalism discussed in~\cite{freire19} to propose a change of variables that produces an experimental independent expression for the irrealism. In the following, we propose a useful way of rescaling the irrealism measure in comparison with the minimum value associated with the experimental resolutions that limit the extraction of irrealism, this is, in terms of the fundamental uncertainty of the system. We consider
$p_{m}\approx\rho(m,t,\tau,\gamma)\delta q$, and we rewrite the
irrealism in Eq. (\ref{Irx}) as
\begin{eqnarray}
\bar{\mathfrak{I}}_{(\delta q\leq \delta
q^{\prime})}(Q|\rho)&\equiv&\mathfrak{I}_{(\delta
q)}(Q|\rho)-\ln(\delta q^{\prime}/\delta q)\nonumber\\
&=&-\delta
q\sum_{m=-N}^{N}\rho(m,t)\ln\left[\rho(m,t)\delta q\right]\nonumber\\
&-&\ln(\delta q^{\prime}/\delta q), \label{Ires}
\end{eqnarray}
where $\Delta x$ is the integration interval for which
the probability is normalized to $1$ and the resolution is related
with $N$ and $\Delta x$ by $\delta q=\Delta x/N$. We can obtain
a similar expression for the momentum irrealism.

\begin{figure}[htp]
\includegraphics[width=4.2 cm]{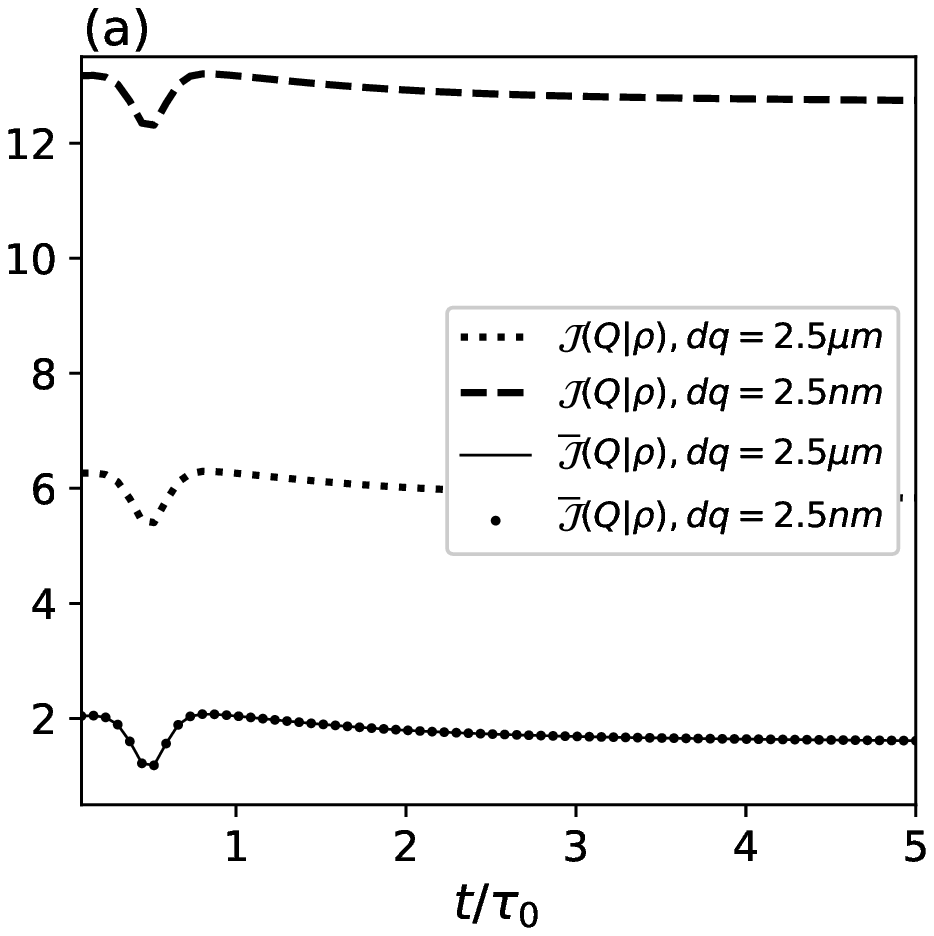}
\includegraphics[width=4.2 cm]{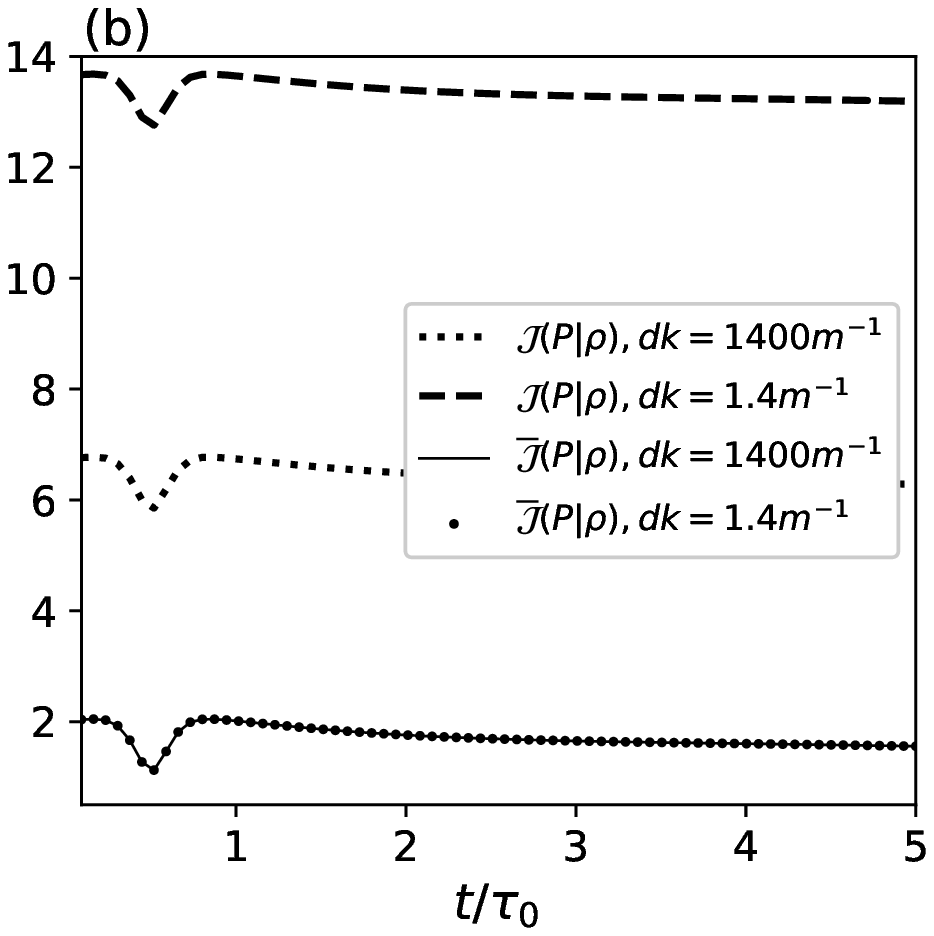}
\caption{Irrealism (a) in position and (b) in momentum as a function
of $t/\tau_{0}$ for $\tau=18\tau_{0}$ and $\gamma=-1.0$.  We
consider two experimental resolutions: $\delta q=2.5\;\mathrm{\mu m}$ (dotted line)
and $\delta q=2.5\;\mathrm{n m}$ (dashed line) for the irrealism in position, and
$\delta k=1.4\;\mathrm{m^{-1}}$ (dashed line) and $\delta k=1400\;\mathrm{m^{-1}}$ (dotted line)
for the irrealism in momentum. These quantities are minima for
$t_{\textsl{min}}\approx0.49\tau_{0}$ and maxima for
$t^{\prime}_{\textsl{max}}\approx0.83\tau_{0}$. In both plots, we compare the irrealism and the rescaled version (solid and full circle lines) to show that the same behavior is preserved.} \label{IrQ}
\end{figure}

Different from the fundamental uncertainties, the irrealism in position and momentum variables
needs to be calculated numerically since we do not have closed
expressions for this particular state. Therefore, we perform numerical calculations and exhibit
them in Fig.\ref{IrQ}. In Fig.\ref{IrQ}(a) we show the plot of the
irrealism in position and in Fig.\ref{IrQ}(b) the irrealism in
momentum as a function of $t/\tau_{0}$ for $\tau=18\tau_{0}$ and
$\gamma=-1.0$. We consider $\delta
q^{\prime}=\sigma_{xx}^{min}\approx0.17\;\mathrm{mm}$ and $\delta
k^{\prime}=\sigma_{pp}^{min}\approx1.58\times10^{5}\;\mathrm{m^{-1}}$
as the experimental resolution of reference. Then, we use two
experimental resolutions $\delta q=2.5\;\mathrm{\mu m}$ and $\delta
q=2.5\;\mathrm{n m}$ for the irrealism in position, and $\delta
k=1.4\;\mathrm{m^{-1}}$ and $\delta k=1400\;\mathrm{m^{-1}}$ for the
irrealism in momentum, which show that the rescaled irrealism of Eq.(\ref{Ires}) 
is resolution-independent. In the same plot we exhibit
the irrealism for these two experimental resolutions, which show that
we have more irrealism with the increasing use of higher-resolution apparatuses. From these results, we see that the relevant physical content is clearly related to the temporal evolution of this quantity. Furthermore, with the rescaling irrealism proposed here, experimentalists could directly use their data, without having to make changes of variables and without bothering with the limit of a higher-resolutions apparatus, since the most important aspect of this model is to employ an apparatus that has more resolution than the fundamental one, which is attached with the uncertainty principle.
The irrealism for each resolution is minimum at the same time for which the fundamental uncertainties and position-momentum correlations are minima, i.e.,
$t_{\textsl{min}}\approx0.49\tau_{0}$. However, the time for the maximum
irrealism is $t^{\prime}_{\textsl{max}}\approx0.83\tau_{0}$ which does not
coincide with the time of maxima position-momentum correlations.

\begin{figure}[htp]
\centering
\includegraphics[width=4.5 cm]{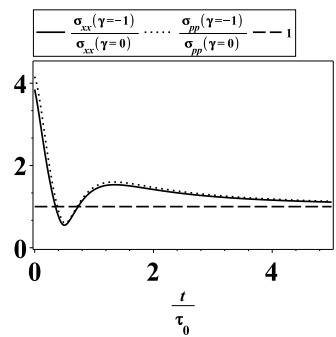}
\caption{Uncertainty in position and in momentum for the state at
the detection screen with $\gamma=-1.0$ as a function of
$t/\tau_{0}$ for $\tau=18\tau_{0}$. The curves are normalized by the
uncertainties for the standard Gaussian superposition  $\gamma=0$.
There is squeezing around the point of minimum
$t_{\textsl{min}}\approx0.49\tau_{0}$.} \label{squeezingxp}
\end{figure}

Note that, by employing such a normalization procedure, we are not claiming that this version is more fundamental than the previous one with the explicit dependence on its experimental resolution. The normalization introduced here could be important for comparing more directly the results obtained previously with different experimental resolutions for the double-slit interferometer setups with matter waves. Before that, it is interesting to discuss the existence of a
point of minimum. Following Bohm in Ref. \cite{Bohm}, the
position-momentum correlations are correlations that develop with
the quantum dynamics producing a minimum Robertson-Schr\"odinger
uncertainty relation for a Gaussian wavepacket. Also, in Bohm's
words, large correlations mean that a high momentum tends to become
correlated with the covering of a large distance. In this case,
minima correlations imply minimum position and momentum
uncertainties that corroborate our results. We can also
understand the existence of minima fundamental and entropic
uncertainties from the viewpoint of squeezing. It has been shown that
the initial contractive state produces a squeezed superposition in
comparison with the standard Gaussian superposition at the detection
screen of a double-slit experiment \cite{IGP2020}. 
In Fig.\ref{squeezingxp}
we show the uncertainties in position and in
momentum for the superposition at the detection screen when
$\gamma=-1.0$, normalized by the respective uncertainties for the
standard Gaussian superposition, i.e., when $\gamma=0$. We can
observe squeezing in position and momentum around the point of
minimum $t_{\textsl{min}}\approx0.49\tau_{0}$.

Therefore, the minimum observed here for the fundamental
uncertainties and irrealism is a consequence
of squeezing present in the superposition when the initial state is
a contractive state. This result presents a similar
behavior to the result obtained in Fig.14 of Ref.\cite{Coles} for
the EPR state \cite{Weedbrook}, which
shows a decreasing in the entropic uncertainty as a function of the
squeezing strength. Squeezing also means a large region of overlap
between the two packets for the time for which the uncertainties and
irrealism are minima. Therefore, when the superposition created in
the double-slit is squeezed/localized in the detection screen to
produce minimal uncertainties and a large region of overlap, which
means that one can not distinguish the packets in the superposition,
the irrealism tends to be minimum. These results are also reflected
in the interference pattern, as we can observe in the next section.

\section{Irrealism and fringe visibility}

It has been shown that wave-particle duality relations are actually
entropic uncertainty in disguise. Also, an
expression  has been found that involves minimum and maximum entropies and unifies
the wave-particle duality principle with the entropic uncertainty
principle; see Eq.(344) of Ref.\cite{Coles}. Here, we calculate
the relative intensity, visibility, and predictability to analyze the
interference pattern as well as the wave and particle properties
from the knowledge of the minimum and maximum irrealism. The minimum entropy, i.e., minimum irrealism, is
characterized by a maximum number of interference fringes and
maximum visibility. The maximum irrealism is characterized by a
small number of interference fringes and small visibility. We also
study the interference pattern, the visibility, and predictability
around the time for which the fundamental uncertainties and
position-momentum correlations are maximal.

\begin{figure}[htp]
\centering
\includegraphics[width=4.25cm]{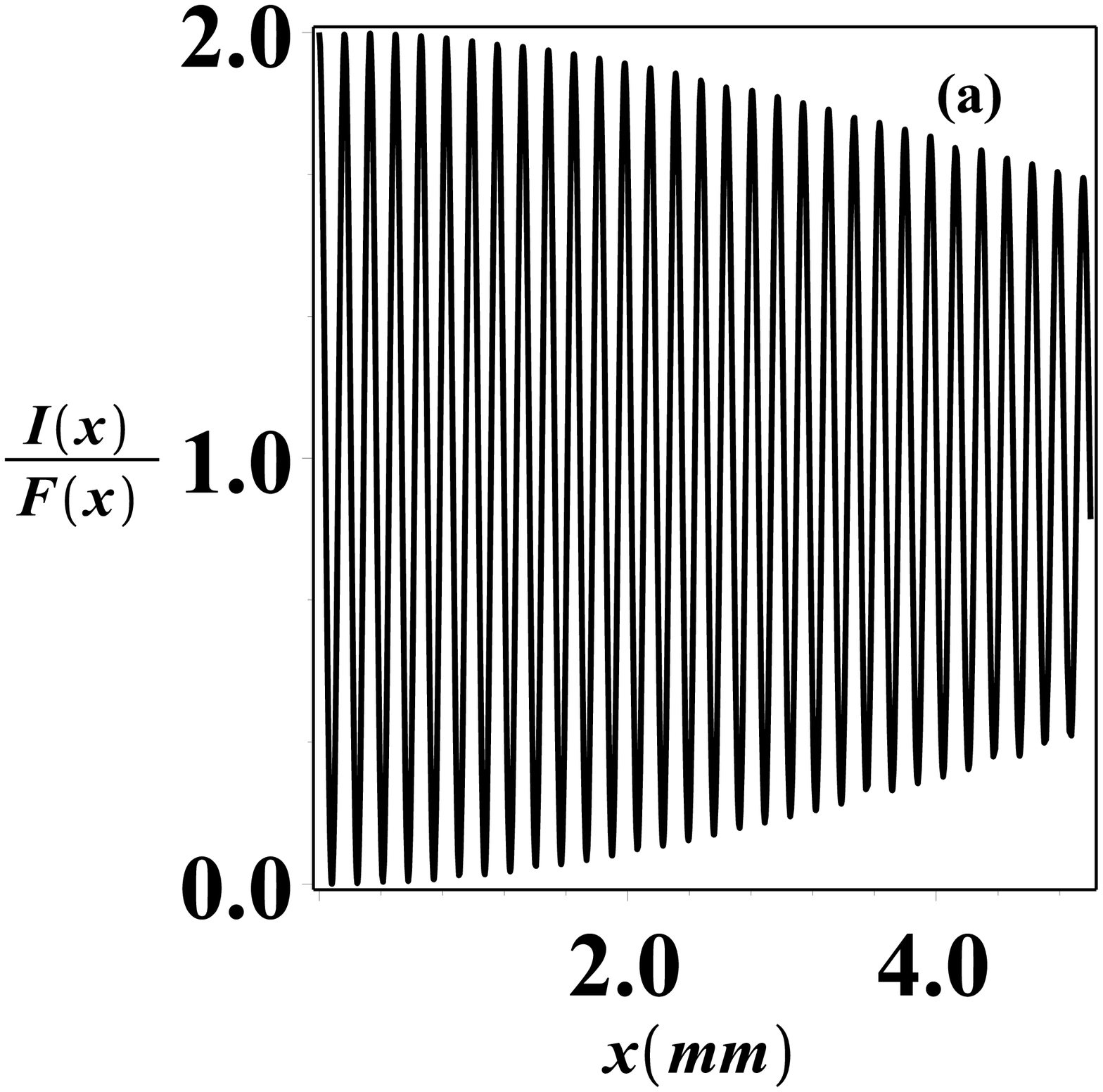}
\includegraphics[width=4.25cm]{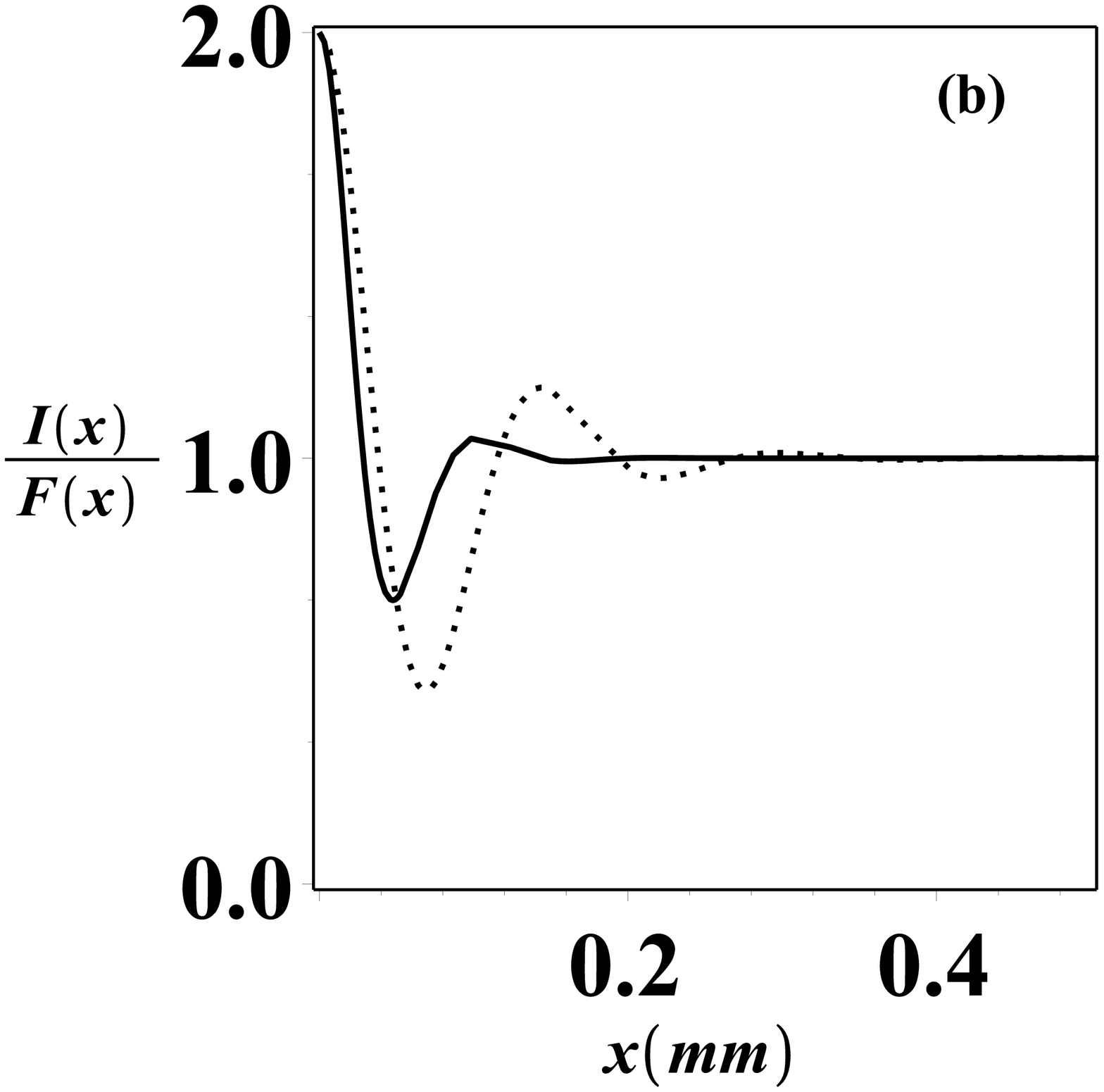}
\caption{Half of the symmetrical plot for the relative intensity as
functions of $x$. (a) The propagation time from the source to the
double-slit is that for which the fundamental and entropic
uncertainties, as well as the irrealism, are minima, i.e.,
$t_{\textsl{min}}\approx0.49\tau_{0}$. (b) The propagation time from the
source to the double-slit is that for which the fundamental
uncertainties and correlations are maxima, i.e.,
$t_{\textsl{max}}\approx1.36\tau_{0}$ (solid line), and for which the
irrealism is maximum, $t_{\textsl{max}}\approx0.83\tau_{0}$ (dotted line).
For the propagation from the double-slit to the screen, we use
$\tau=18\tau_{0}$.} \label{int_rel}
\end{figure}

The intensity on the screen is given by
\begin{equation}
I(x,t,\tau)=F(x,t,\tau)+F(x,t,\tau)\frac{\cos(2\Delta_{\gamma}
x)}{\cosh(\frac{D_{\gamma} x}{B_{\gamma}^{2}})}\label{intensidade1},
\end{equation}
where
\begin{equation}
F(x,t,\tau)=\frac{2}{B_{\gamma}\sqrt{\pi}}\exp\left[-\frac{x^{2}+(\frac{D_{\gamma}}{2})^{2}}{B_{\gamma}^{2}}\right]\cosh\left(\frac{D_{\gamma}
x}{B_{\gamma}^{2}}\right).
\end{equation}
The first term in Eq.(\ref{intensidade1}) is the single-slit
envelope, and the second term is the interference \cite{solano}. From
Eq.(\ref{intensidade1}), we obtain the expression for the relative
intensity,
\begin{equation}
\frac{I(x,t,\tau)}{F(x,t,\tau)}=1+\frac{\cos(2\Delta_{\gamma}
x)}{\cosh(\frac{D_{\gamma} x}{B_{\gamma}^{2}})}\label{intensidade2}.
\end{equation}
In Fig.\ref{int_rel}, we show half of the symmetrical plot for the
relative intensity as functions of $x$ for an initial contractive
state. In Fig. \ref{int_rel}(a) we consider the time for which the
irrealism is minimum, i.e., $t_{\textsl{min}}\approx0.49\tau_{0}$. In Fig.
\ref{int_rel}(b) we consider the time for which the irrealism is
maximum, i.e., $t^{\prime}_{\textsl{max}}\approx0.83\tau_{0}$ (dotted line)
and the time for which the fundamental uncertainties and
position-momentum correlations are maximal, i.e.,
$t_{\textsl{max}}\approx1.36\tau_{0}$ (solid line). We fixed the propagation
time from the double-slit to the screen in $\tau=18\tau_{0}$. We
observe a large number of interference fringes associated with the
minimum irrealism and a small number of interference fringes
associated with the maximum irrealism. A smaller number of
interference fringes are observed for the time of maxima fundamental
uncertainties and position-momentum correlations.

The knowledge of both ``particle" and ``wave" in an interferometric
experiment is given by the Greenberger and Yasin expression
$\mathcal{P}(\theta)^{2}+\mathcal{V}(\theta)^{2}\leq1$, where
$\mathcal{P}$ stands for particle property and $\mathcal{V}$ for
wave property. The parameter ($\theta$) is used to vary from full
particle to full wave knowledge preserving the general case in which
one can have considerable knowledge of both. The equality is ensured
for pure quantum-mechanical states and the inequality for mixed
states \cite{Greenberger}. We calculate the predictability and
visibility for our experimental setup and obtain
\begin{equation}
\mathcal{P}(x)=\left|\frac{|\psi_{1}|^{2}-|\psi_{2}|^{2}}{|\psi_{1}|^{2}+|\psi_{2}|^{2}}\right|=\left|\tanh\left(\frac{D_{\gamma}x}{B_{\gamma}^{2}}\right)\right|,
\end{equation}
and
\begin{equation}
\mathcal{V}(x)=\frac{I_{\textsl{max}}-I_{\textsl{min}}}{I_{\textsl{max}}+I_{\textsl{min}}}=\frac{1}{\cosh(\frac{D_{\gamma}x}{B_{\gamma}^{2}})},
\label{vis}
\end{equation}
where $I_{\textsl{max}}$ is the intensity for $\cos(2\Delta_{\gamma} x)=1$
and $I_{\textsl{min}}$ is the intensity for $\cos(2\Delta_{\gamma} x)=-1$
\cite{solano}. Similar results were obtained previously in Ref.
\cite{Bramon}.

\begin{figure}[htp]
\centering
\includegraphics[width=4.25 cm]{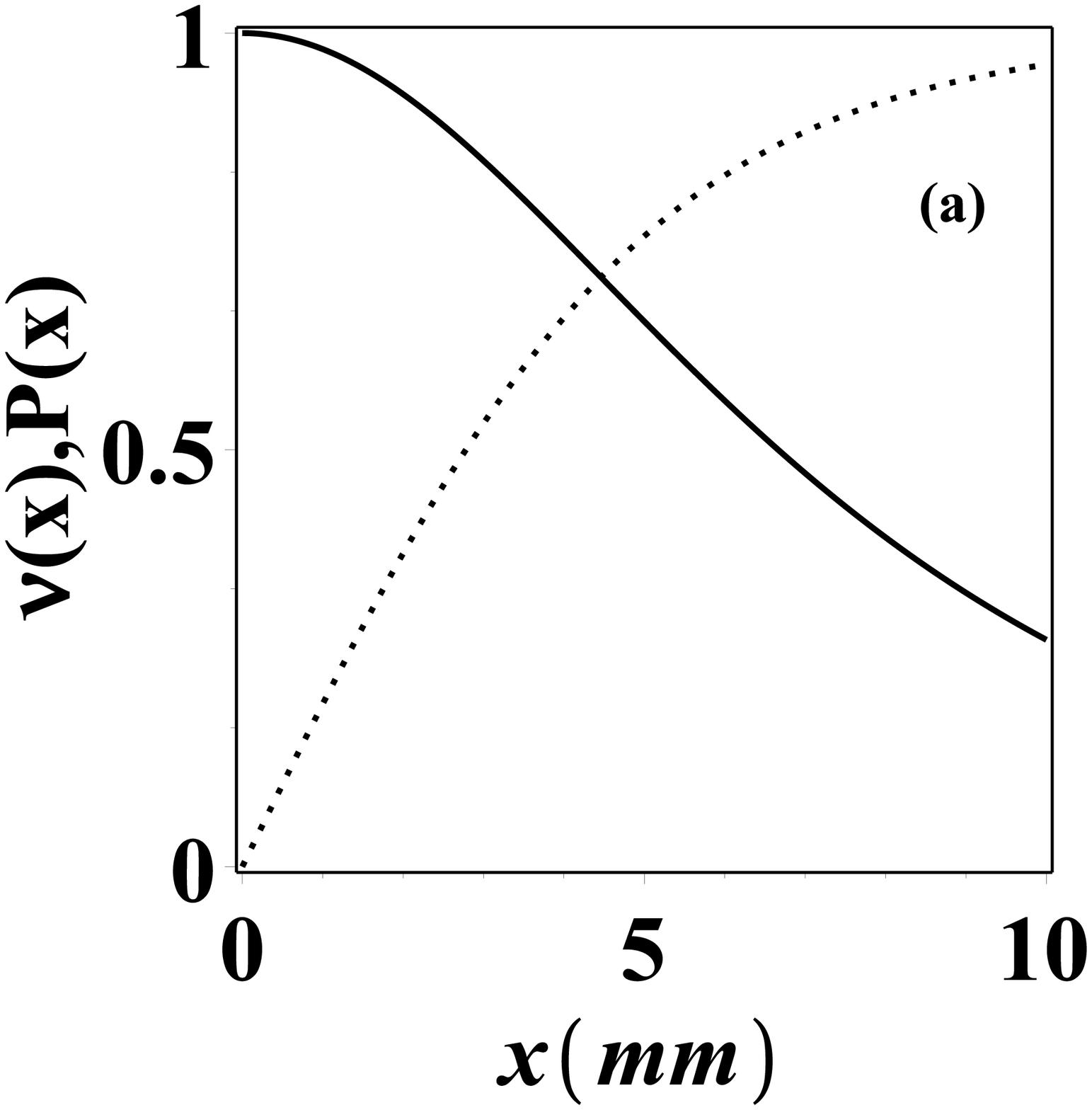}
\includegraphics[width=4.25 cm]{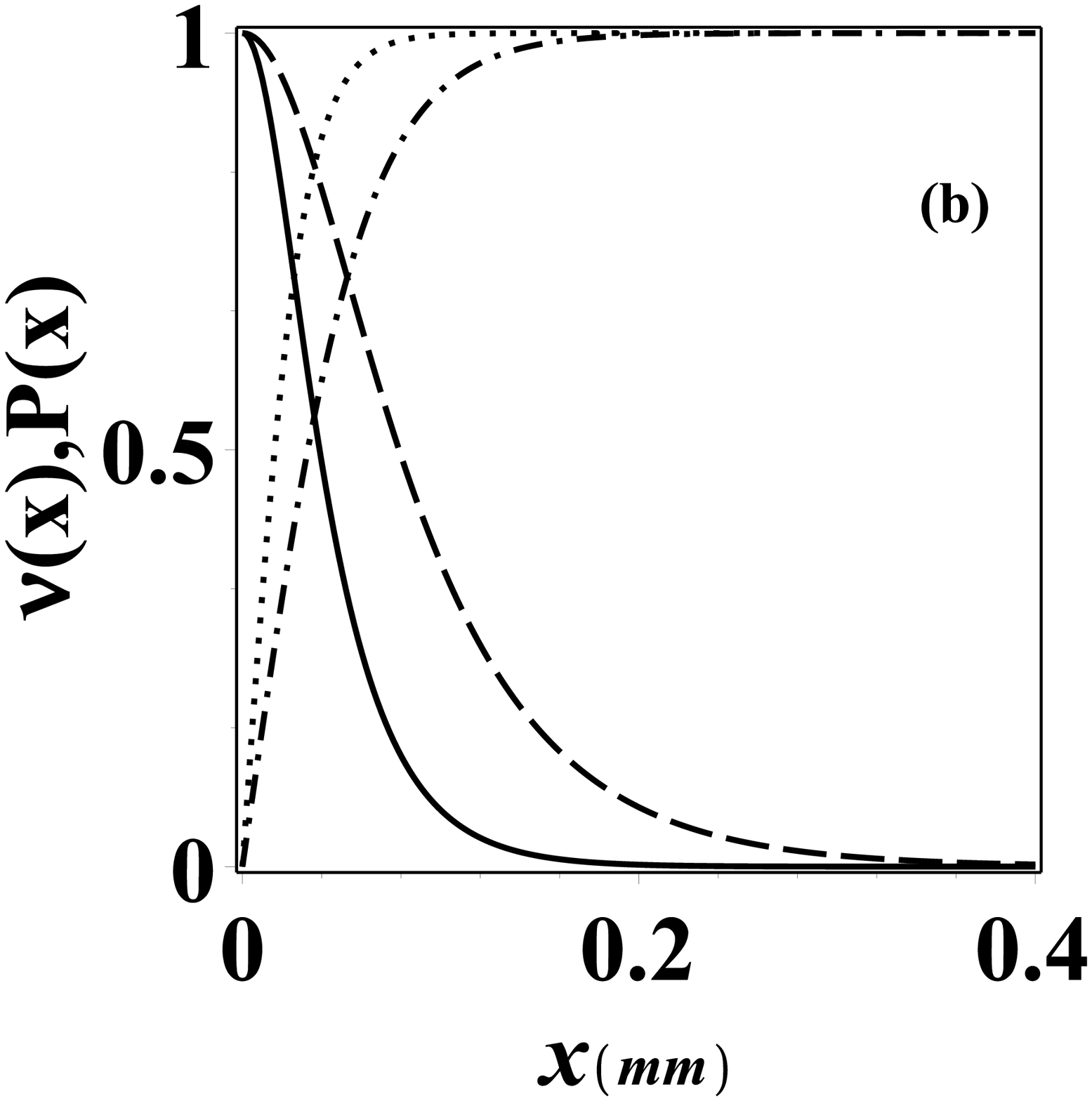}
\caption{Half of the symmetrical plot of the visibility
and predictability as functions of $x$. (a) Time for
which the fundamental and entropic uncertainties as well as the
irrealism are minimal, and (b) time for which the fundamental
uncertainties and the position-momentum correlations are maxima
(solid and dotted lines) and for which the irrealism is maximum (dashed and dash-dotted lines).
$\tau=18\tau_{0}$ for both plots.} \label{vis_prev}
\end{figure}

In Fig.\ref{vis_prev}(a) we show half of the symmetrical plot of
the visibility (solid line) and predictability (dotted line) as
functions of $x$ for the time for which the fundamental and entropic
uncertainties, as well as the irrealism, are minimal. In Fig.
\ref{vis_prev}(b) we show half of the symmetrical plot of the
visibility (solid line) and predictability (dotted line) as
functions of $x$ for the time for which the fundamental
uncertainties and the position-momentum correlations are maximal,
 and the plot of the visibility (dashed line) and
predictability (dash-dotted line) for the time for which the irrealism is
maximum. As before, we fixed the propagation time from
the double-slit to the screen in $\tau=18\tau_{0}$.

These results show clearly the relationship between the minimum of
the entropic uncertainty and irrealism with the maximum number of
interference fringes and visibility (wave character) in the
double-slit experiment. The wave property is predominant in a larger
region of the axis $x$ when the irrealism is minimum and the
particle property is dominant around the maximum irrealism, as we can
observe by comparing the curves of visibility and predictability for
the times of minimum and maximum irrealism. These results are
according to the connection obtained in Ref.\cite{Coles}, which
associates the wave-particle duality with minimum and maximum
entropy.

\begin{figure}[htp]
\centering
\includegraphics[width=4.2 cm]{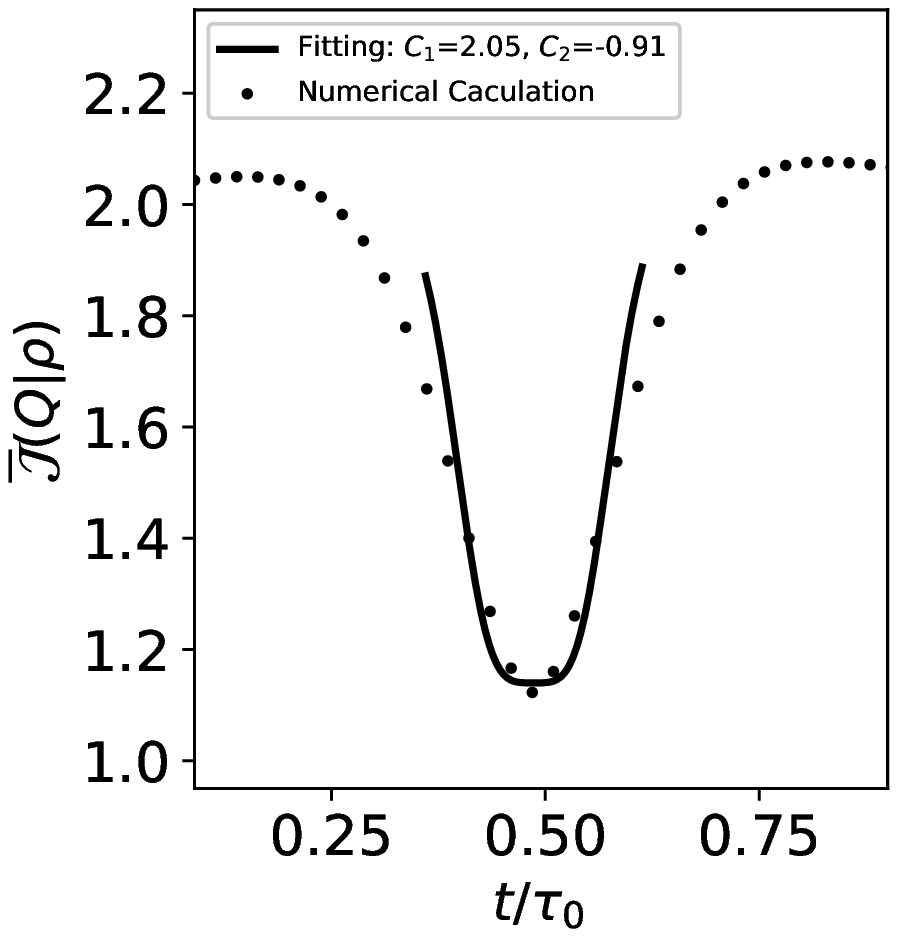}
\includegraphics[width=4.2 cm]{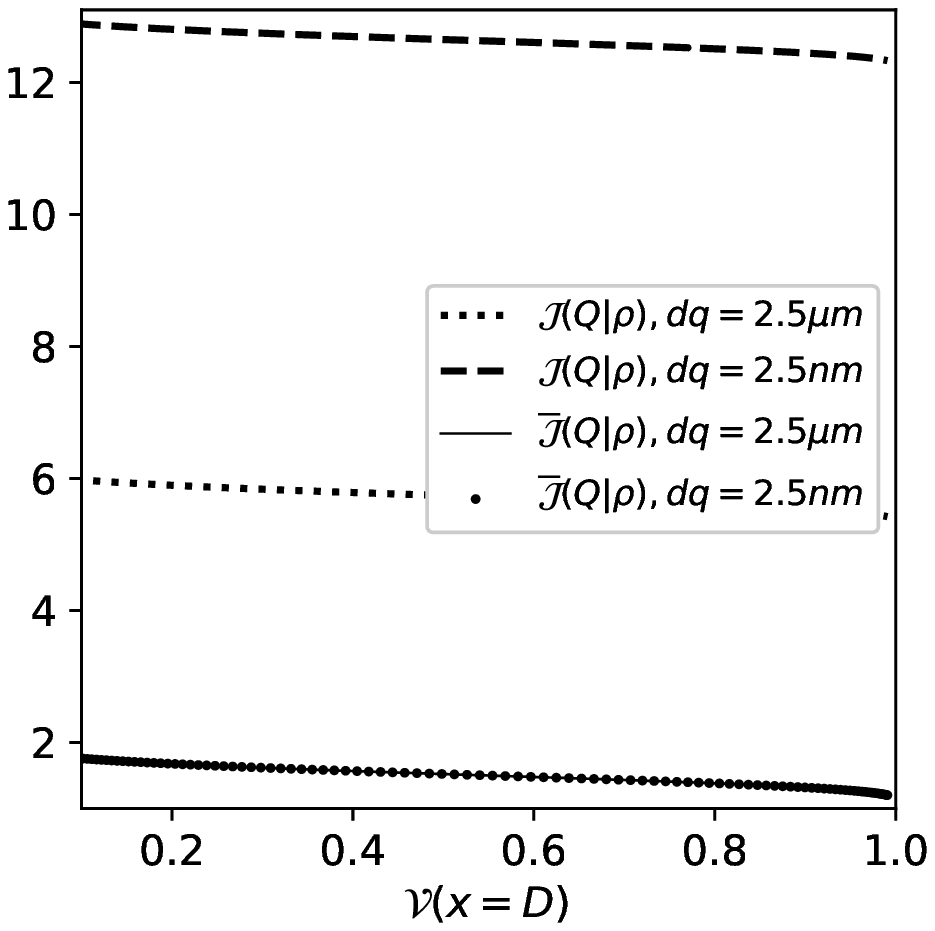}
\caption{In the left figure, we plot the normalized irrealism in position as a function of $t/\tau_{0}$. The full circle
curve is the numerical calculation and the solid curve is the adjustment
by Eq. (\ref{adjust}). In the right figure is
the irrealism as a function of the visibility.} \label{irrealism_vis}
\end{figure}

The results above suggested that for the point of minimum the
irrealism can be obtained from the visibility. We find that around
this point the irrealism can be approximated by the equation
\begin{eqnarray}
\bar{\mathfrak{I}}(Q|\rho)\approx
C_1+C_2\mathcal{V}[x=D_{\gamma}(t\cong t_{\textsl{min}})], \label{adjust}
\end{eqnarray}
where $C_1$ and $C_2$ are constants to be found in the adjustment, and
$\mathcal{V}[x=D_{\gamma}(t\cong t_{\textsl{min}})]$ is the fringe visibility
measured in the position $x$ equal to the wave-packet separation
$D_{\gamma}(t\cong t_{\textsl{min}})$ while the source is placed in the
position $z=v_z(t\cong t_{\textsl{min}})$. Then, each value of $t\cong
t_{\textsl{min}}$ which gives a specific value for $z$ and $x$ produces a
specific value for the visibility and irrealism. 
In Fig.\ref{irrealism_vis} we show the plot for the irrealism in position
as a function of $t/\tau_{0}$. The full circle curve is the numerical
calculation, and solid curve is the adjustment Eq. (\ref{adjust}). In
adjusting we find $C_1=2.05$ and $C_2=-0.91$. We can observe from
Eq. (\ref{adjust}) and Fig.\ref{irrealism_vis} that the minimum
irrealism is related with the maximum visibility. Furthermore, as we can see in the parametric plot, these results show a monotonic relation between irrealism and visibility in the right plot of Fig. \ref{irrealism_vis}. As expected, we observe the same behavior for the irrealism and the rescaled one.

Therefore, around the point of minimum, where the visibility is
large, one can measure the fringe visibility as a function of
$D_{\gamma}(t\cong t_{min})$ and use Eq. (\ref{adjust}) to obtain
the corresponding measurement of irrealism.

\section{Conclusions}

The purely epistemic uncertainty present in classical statistical
physics does not reflect the irrealism of the
physical properties of these systems. In addition to our
statistical ignorance,  these properties are already predetermined
before any measurement in such a classical regime; in this case, the uncertainty reflects only subjective ignorance. On the other hand,
quantum superposition, which lies at the heart of the quantum theory,
points out that reality seems to be in suspension in such cases,
that is, physical properties do not have well-defined values that
support physical realism. This gives rise to many of the contra-intuitive interpretations of quantum theory since even molecules exhibit this condition. This ontological uncertainty implies that the classical notion of realism is
forbidden in general for two noncommuting observables such as
position and momentum, as demonstrated in~\cite{freire19}.

In this work, employing the quantifier introduced in~\cite{freire19}, and proposing a rescaling in terms of the fundamental uncertainties, we have taken a step further on this issue by
exploring the consequences of an initial contractive state to the degree of position and momentum irrealism in double-slit configuration for massive particles interference. We analyzed the connection of physical irrealism and fundamental uncertainties, with the
intensity, visibility, and predictability of the wavepackets
interference. We saw that the propagation time that minimizes the fundamental uncertainties and the irrealism in position or momentum are the
same. This minimum is related to the squeezing present in the
superposition for an initial contractive state, since a non-contractive state has only a maximum value.
By analyzing the behavior of the visibility and predictability for
the minimum and maximum times, we saw that minimum irrealism
coincides with the maximum number of interference fringes and
visibility for this configuration, while predictability is
dominant in the region of maximum irrealism. Moreover, this coincidence in the minimum
indicates that it is possible to estimate irrealism if we look at
visibility at this same point of the minimum. Indeed, we proposed here a
way to indirect probe irrealism in the context of matter-wave
double-slit interference by probing the fringe visibility.

Finally, the results of this work are important to the proposal of a
successful interference experiment, indicating how to define values
for parameters that can produce a maximum number of fringes with
better visibility. This is also important from the perspective of
the foundations of quantum theory, indicating that there is a
minimum value for the irrealism of the studied system that produces
the maximum interference in the context of matter waves, which
allows us to reinterpret the double-slit experiment with the initial contractive state by employing a notion of a state with fundamental physical indefiniteness, instead
of thinking of a particle traveling as a definite wave. This highlights the quantum nature of this mechanism of interference and the lack of reality of that systems.
We hope that the results developed here help to encourage experimentalists to implement an indirect
measurement of irrealism to matter waves.

\vskip1.0cm
\begin{acknowledgments}
The authors would like to thank the anonymous referees for their many insightful comments and suggestions. The authors acknowledge CAPES and CNPq-Brazil for financial
support. P.R.D. acknowledges Grant No. 88887.354951/2019-00 from
CAPES. I. G. da Paz acknowledges Grant No. 307942/2019-8 from CNPq.
\end{acknowledgments}


\end{document}